\documentclass[twocolumn,sort, compress, 10pt, letterpaper]{svjour3}
\usepackage[latin9]{inputenc}
\usepackage{array}
\usepackage{float}
\usepackage{multirow}
\usepackage{amsmath}
\usepackage{amssymb}
\usepackage{graphicx}
\usepackage[numbers]{natbib}

\makeatletter

\providecommand{\tabularnewline}{\\}
\floatstyle{ruled}
\newfloat{algorithm}{tbp}{loa}
\providecommand{\algorithmname}{Algorithm}
\floatname{algorithm}{\protect\algorithmname}

\RequirePackage{fix-cm}
\smartqed  
\DeclareGraphicsExtensions{.pdf,.eps,.jpg}
\usepackage{algorithmic}
\usepackage{algorithm}
\usepackage{epstopdf}

\makeatother

\begin{document}
\title{
Computer Vision Accelerators for Mobile Systems based on OpenCL GPGPU Co-Processing
}


\author{Guohui Wang	\and
        Yingen Xiong   \and
		Jay Yun		\and
		Joseph R. Cavallaro
}


\institute{G. Wang and J. R. Cavallaro \at
		   Department of Electrical and Computing Engineering \\
		   Rice University, Houston, Texas-77005, USA\\
              \email{wgh@rice.edu, cavallar@rice.edu}           
           \and
           Y. Xiong and J. Yun \at
           Qualcomm Technologies Inc., San Diego, California, USA
}

\date{Received: / Accepted: }
\titlerunning{J Sign Process Syst (2014)}
\authorrunning{Guohui Wang et al.}

\maketitle
\begin{abstract}
In this paper, we present an OpenCL-based heterogeneous implementation
of a computer vision algorithm -- image inpainting-based object removal
algorithm -- on mobile devices. To take advantage of the computation
power of the mobile processor, the algorithm workflow is partitioned
between the CPU and the GPU based on the profiling results on mobile
devices, so that the computationally-intensive kernels are accelerated
by the mobile GPGPU (general-purpose computing using graphics processing
units). By exploring the implementation trade-offs and utilizing the
proposed optimization strategies at different levels including algorithm
optimization, parallelism optimization, and memory access optimization,
we significantly speed up the algorithm with the CPU-GPU heterogeneous
implementation, while preserving the quality of the output images.
Experimental results show that heterogeneous computing based on GPGPU
co-processing can significantly speed up the computer vision algorithms
and makes them practical on real-world mobile devices. 
\end{abstract}

\section{Introduction}

Mobile devices have evolved rapidly and become ubiquitous over the
past decade, giving rise to new application demands through the convergence
of mobile computing, wireless communication and digital imaging technologies.
On one hand, as mobile processors have become more and more powerful
and versatile during the past several years, we are witnessing a rapid
growth in the demand for the computer vision applications running
on mobile devices, such as image editing, augmented reality, object
recognition and so on~\citep{wagner:ISMAR2008:mobile_pose_tracking,Lee:ICCD2009:mobile_AR,paucher:CVPR2010:location_AR_mobile,fernandez:CMMSE2012:performance_mobile_AR,yang:ISMAR2012:ldb,Yang_ACMMM2012:mobile_surf,Clemons:2012:arch_mobile_vision,Pulli_ACM2012:OpenCV}.
On the other hand, with the recent advances in the fields of computer
vision and augmented reality, the emerging algorithms have become
more complex and computationally-intensive. Therefore, the long processing
time due to the high computational complexity prevents these computer
vision algorithms from being practically used in mobile applications. 

To address this problem, researchers have been exploring general-purpose
computing using graphics processing units (GPGPUs) as accelerators
to speed up the image processing and computer vision algorithms thanks
to the heterogeneous architecture of the modern mobile processors~\citep{Leskela_NOKIA_SIPS_2009,Singhal_ICIP2010:mobile_image_processing,ensor:arXiv2011:mobile_gpu_image_analysis,bordallo:SPIE2011:image_reg_mobile_GPGPU,Cheng_VLSIDAT2011:face_recognition,hofmann:ISMAR2012:gpgpu_descriptor,wang:SIPS2012:energy_mobile,hofmann:thesis2012:extraction_mobile_gpu,ICASSP2013_Rister_SIFT}.
On desktop computers or supercomputers, numerous programming models
have been extensively studied and utilized to facilitate the parallel
GPGPU programming, such as the Compute Unified Device Architecture
(CUDA)~\citep{cuda} and the Open Computing Language (OpenCL)~\citep{opencl,opencl_programming_guide}.
As a comparison, due to the lack of parallel programming models in
the mobile domain, the OpenGL ES (Embedded System) programming model
was commonly used to harness the computing power of the mobile GPU~\citep{opengl_es}.
However, the inherent limitations of the OpenGL ES lead to poor flexibility
and scalability, as well as limited parallel performance, due to the
fact that the OpenGL ES was original designed for 3D graphics rendering.
Recently, emerging programming models such as the OpenCL embedded
profile~\citep{opencl} and the RenderScript~\citep{android} have
been supported by the state-of-the-art mobile processors, making the
mobile GPGPU feasible for real-world mobile devices for the first
time~\citep{Pulli_ACM2012:OpenCV,ICASSP2013_Wang,Wang:GlobalSIP2013:SIFT_GPU}. 

In this paper, we take the exemplar-based image inpainting algorithm
for object removal as a case study to explore the capability of the
mobile GPGPU to accelerate computer vision algorithms using OpenCL.
The remainder of this paper is organized as follows. Section \ref{sec:GPGPU-on-Mobile}
introduces the architecture of the modern mobile SoCs and the OpenCL
programming model for the mobile GPGPU. Section \ref{sec:algorithm-overview}
briefly explains the exemplar-based inpainting algorithm for object
removal. We analyze the algorithm workflow and propose a method to
map the algorithm between mobile CPU and GPU in Section \ref{sec:Workflow-Analysis}.
To adapt the complex algorithm to the limited hardware resources on
mobile processors, we further study implementation trade-offs and
optimization strategies to reduce the processing time in Section~\ref{sec:Algorithm-Optimizations}.
Section \ref{sec:Experimental-Results} shows experimental results
on a practical mobile device, which indicates that our optimized GPU
implementation shows significant speedup, enabling fast interactive
object removal applications in a practical mobile device. Section
\ref{sec:Conclusions} concludes the paper.

\section{GPGPU on Mobile Devices\label{sec:GPGPU-on-Mobile}}

\begin{figure}
\begin{centering}
\includegraphics[width=0.8\columnwidth]{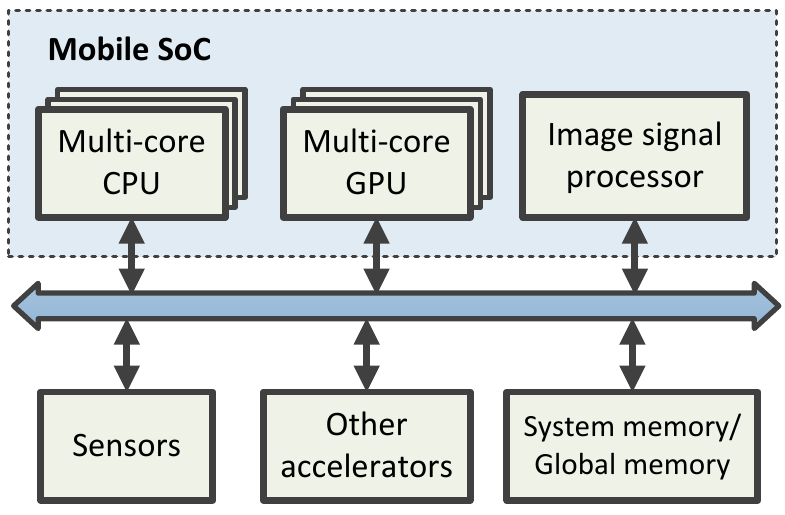}
\par\end{centering}

\caption{Architecture of a typical mobile platform.\label{fig:mobile_SoC}}
\end{figure}

As is shown in Fig.~\ref{fig:mobile_SoC}, a modern mobile SoC (system-on-chip)
chipset typically consists of a multi-core mobile CPU, a mobile GPU
with multiple programmable shaders, and an image signal processor
(ISP)~\citep{Snapdragon,PowerVR,ULP_GeForce}. Unlike their desktop
counterparts, the mobile CPU and GPU share the same system memory
via a system bus. The mobile platforms also contain a variety of sensors
and accelerators. Modern mobile platforms tend to employ heterogeneous
architectures, which integrate several application-specific co-processors
to enable the computationally intensive algorithms such as face detection
and so on. However, the space limitation and the power constraints
of the mobile devices limit the number of integrated hardware co-processors.
It is preferable to seek the general-purpose computing power inside
the mobile processor. The mobile GPUs are suitable candidates to accelerate
computationally intensive tasks due to their highly parallel architecture. 

\begin{figure}
\begin{centering}
\includegraphics[width=0.8\columnwidth]{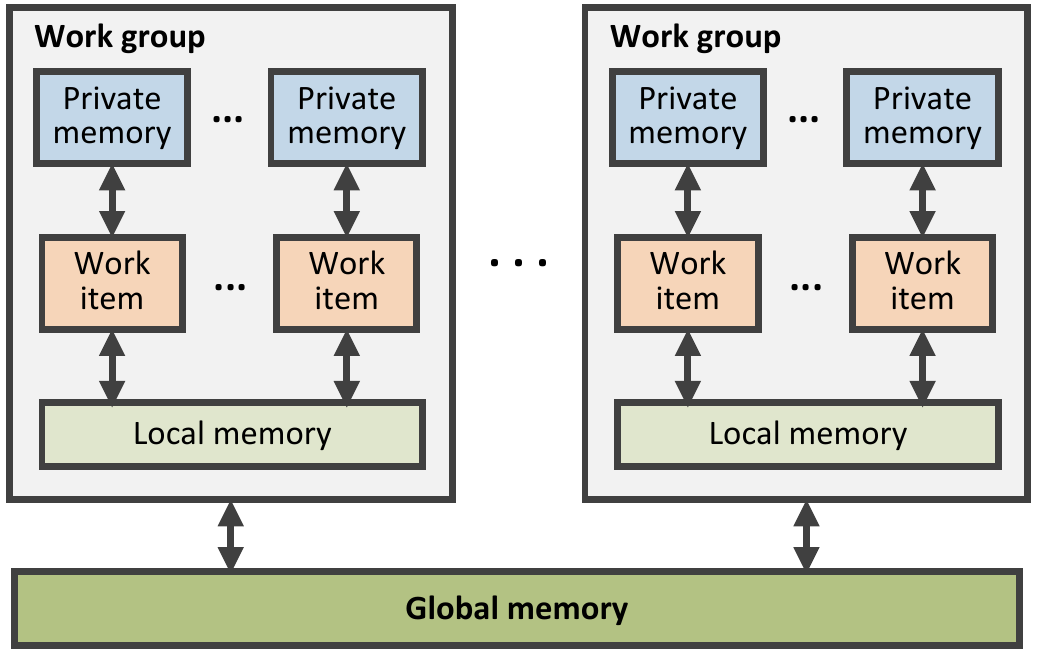}
\par\end{centering}

\caption{OpenCL programming model and hierarchical memory architecture.\label{fig:OpenCL-programming-model}}
\end{figure}

The lack of good parallel programming models becomes an obstacle to
perform general-purpose computing on the mobile GPUs. As a compromise,
researchers have been using the OpenGL ES programming model for GPGPU
to achieve performance improvement and energy efficiency on mobile
devices during the past decade. For instance, Singhal et al. implemented
and optimized an image processing toolkit on handheld GPUs~\citep{Singhal_ICIP2010:mobile_image_processing}.
Nah et al. proposed an OpenGL ES-based implementation of ray tracing,
called MobiRT, and studied the CPU-GPU hybrid architecture~\citep{nah:SIGGRAPH-A2010:mobirt}.
Ensor et al. presented GPU-based image analysis on mobile devices,
in which the Canny edge detection algorithm was implemented using
the OpenGL ES~\citep{ensor:arXiv2011:mobile_gpu_image_analysis}.
Researchers have also attempted to accelerate feature detection and
extraction using the mobile GPUs ~\citep{bordallo:SPIE2011:image_reg_mobile_GPGPU,yang:ISMAR2012:ldb,hofmann:thesis2012:extraction_mobile_gpu,ICASSP2013_Rister_SIFT}.
Recently, performance characterization and energy efficiency for mobile
CPU-GPU heterogeneous computing have been studied~\citep{Cheng_VLSIDAT2011:face_recognition,wang:SIPS2012:energy_mobile}.

\begin{figure*}
\begin{centering}
\includegraphics[width=0.65\textwidth]{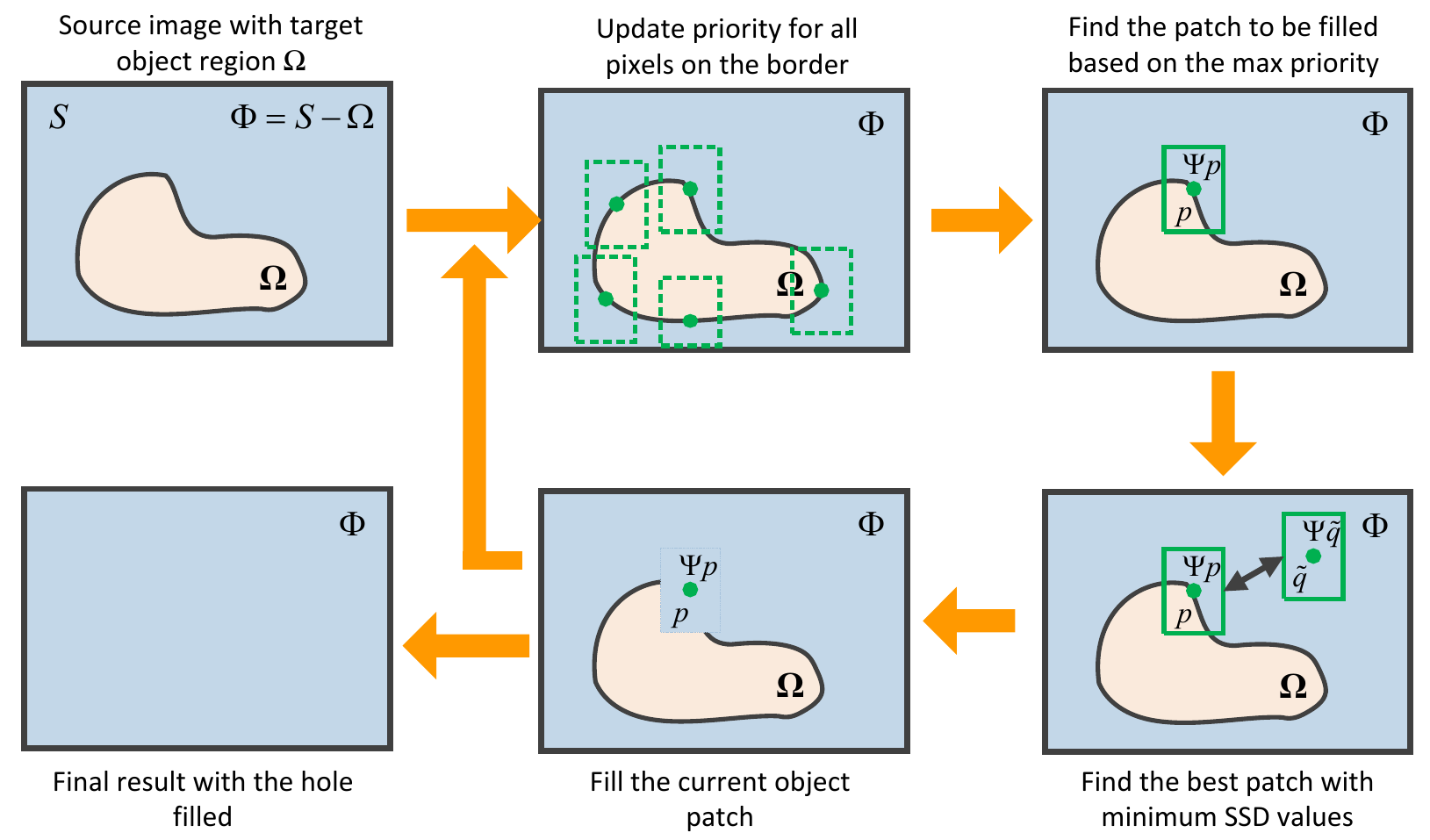}
\par\end{centering}

\caption{Major steps of the exemplar-based image inpainting algorithm for object
removal.\label{fig:algorithm_steps}}
\end{figure*}

Thanks to the unified programmable shader architecture and the emerging
parallel programming frameworks such as OpenCL, the new generation
of mobile GPUs have gained real general-purpose parallel computing
capability. OpenCL is a programming framework designed for heterogeneous
computing across various platforms~\citep{opencl}. Fig.~\ref{fig:OpenCL-programming-model}
shows the programming and the hierarchical memory architecture of
OpenCL. In OpenCL, a host processor (typically a CPU) manages the
OpenCL context and is able to offload parallel tasks to several compute
devices (for instance, GPUs). The parallel jobs can be divided into
work groups, and each of them consists of many work items which are
the basic processing units to execute a kernel in parallel. OpenCL
defines a hierarchical memory model containing a large off-chip global
memory but with long latency of several hundred clock cycles, and
a small but fast on-chip local memory which can be shared by work
items in the same work group. To efficiently and fully utilize the
limited computation resources on a mobile processor to achieve high
performance, partitioning the tasks between CPU and GPU, exploring
the algorithmic parallelism, and optimizing the memory access need
to be carefully considered. Few prior works have studied the methodology
of using OpenCL to program mobile GPUs and achieve speedup. Leskela
et al. demonstrated a prototype of OpenCL Embedded Profile emulated
by OpenGL ES on mobile devices and showed advantages in performance
and energy efficiency~\citep{Leskela_NOKIA_SIPS_2009}. In our previous
work, we have explored the mobile GPGPU capability of mobile processors
to accelerate computer vision algorithms such as an image editing
algorithm (object removal), and feature extraction based on SIFT (scale-invariant
feature transform)~\citep{ICASSP2013_Wang,Wang:GlobalSIP2013:SIFT_GPU}.
Performance improvement and energy consumption reduction have been
observed.

\section{Overview of Exemplar-based Object Removal Algorithm\label{sec:algorithm-overview}}

\begin{figure}
\begin{centering}
\includegraphics[width=1\columnwidth]{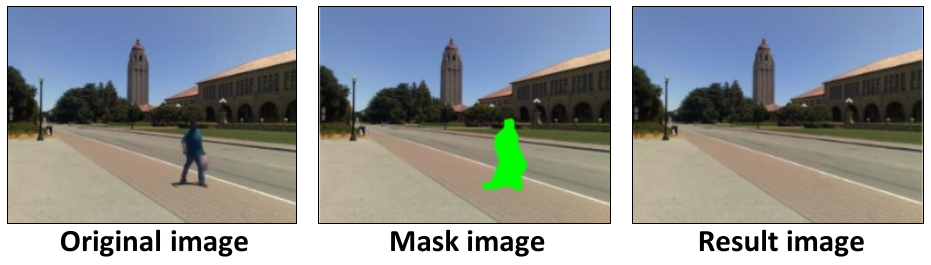}
\par\end{centering}

\caption{An example of the object removal algorithm. The mask image indicates
the object to be removed from the original image.}

\label{fig:idea_of_object_removal}
\end{figure}

In this paper, we take the exemplar-based inpainting algorithm for
object removal as a case study to show the methodology of using the
mobile GPU as a co-processor to accelerate computer vision algorithms.
The object removal algorithm involves raw image pixel manipulation,
iterative image processing technique, sum of squared difference (SSD)
computation and so on, which are typical operations for many computer
vision algorithms. Therefore, the case study of the object removal
implementation can represent a class of computer vision algorithms,
such as image stitching, object recognition, motion estimation, texture
analysis and synthesis, and so on. Therefore, by studying and evaluating
the performance of the exemplar-based object removal algorithm on
mobile devices with CPU-GPU partitioning, the feasibility and advantages
of using the mobile GPU as a co-processor can be demonstrated. Furthermore,
the optimization techniques proposed in this paper can possibly be
applied to other computer vision algorithms with similar operation
patterns or algorithm workflows. 

Object removal is one of the most important image editing functions.
As is shown in Fig.~\ref{fig:idea_of_object_removal}, the key idea
of object removal is to fill in the hole that is left behind after
removing an unwanted object, to generate a visually plausible result
image. The exemplar-based inpainting algorithm for object removal
can preserve both structural and textural information by replicating
patches in a best-first order, which can generate good image quality
for object removal applications~\citep{cvpr2003_Criminisi,asilomar2009_Xiong}.
In the meanwhile, this algorithm can achieve computational efficiency
thanks to the block-based sampling processing, which is especially
attractive for a parallel implementation. 

\begin{table}
\centering{}\footnotesize

\caption{Specification of the experimental setup.\label{tab:experimental_setup}}

\begin{centering}
\begin{tabular}{c|c}
\hline 
\textbf{Mobile SoC} & \textbf{Snapdragon 8974}\tabularnewline
\hline 
\hline 
\textbf{CPU} & \textbf{Krait 400 Quad-core}\tabularnewline
\hline 
Max clock frequency & 2.15 GHz\tabularnewline
\hline 
Compute units & 4\tabularnewline
\hline 
Local memory & 32 KB/compute unit\tabularnewline
\hline 
\hline 
\textbf{GPU} & \textbf{Adreno 330}\tabularnewline
\hline 
Max clock frequency & 400 MHz\tabularnewline
\hline 
Compute units & 4\tabularnewline
\hline 
Local memory & 8 KB/compute unit\tabularnewline
\hline 
\hline 
Operating system & Android 4.2.2\tabularnewline
\hline 
Development toolset & Android NDK r9\tabularnewline
\hline 
Instruction set & ARM-v7a\tabularnewline
\hline 
\end{tabular}
\par\end{centering}

\normalsize
\end{table}

The major steps of the exemplar-based inpainting algorithm for object
removal proposed by Criminisi et al. is depicted in Fig.~\ref{fig:algorithm_steps}~\citep{cvpr2003_Criminisi}.
Assume we have a source image $S$ with a target region $\Omega$
to be filled in after an object is removed (the empty region). The
left image region is denoted as $\Phi$ ($\Phi=S-\Omega$). The border
of the object region is denoted as $\delta\Omega$. The image patches
are filled into the object region $\Omega$ one by one based on priority
values $C(p)$. Given an image patch $\Psi_{p}$ centered at pixel
$p$ for $p\in\delta\Omega$, the priority value $C(p)$ is defined
as the product of two terms:

\begin{equation}
P(p)=R(p)\cdot D(p),
\end{equation}
in which $R(p)$ is the confidence term indicating the amount of reliable
information surrounding the pixel $p$, and $D(p)$ is the data term
representing the strength of texture and structural information along
the edge of the object region $\delta\Omega$ in each iteration. $R(p)$
and $D(p)$ are defined as follows:

\begin{eqnarray}
C(p) & = & \frac{\sum_{q\in\Psi_{p}\cap\Omega}C(q)}{|\Psi_{p}|},\nonumber \\
D(p) & = & \frac{\nabla I_{p}^{\perp}\cdot n_{p}}{\alpha},
\end{eqnarray}
 where $|\Psi_{p}|$ is the area of $\Psi_{p}$, $\alpha$ is a normalization
factor (for a typical grey-level image, $\alpha=255$), and $n_{p}$
is a unit vector orthogonal to $\delta\Omega$ in the point $p$.

According to the priority values for all patches across the border
$\delta\Omega$ of the target region, we select a candidate patch
with the maximum priority value. Then, we search the image region
$\Phi$ and find a patch $\Psi_{\tilde{q}}$ that best matches the
patch $\Psi_{p}$ (we call this step \emph{findBestPatch}). The goal
of \textit{findBestPatch} is to find the best matching patch $\Psi_{\tilde{q}}$
from candidate patches $\Psi_{q}$ in the source image region $\Phi$,
to match an object patch $\Psi_{p}$ in the object region $\Omega$
based on a certain distance metric. The sum of squared differences
(SSD) is typically used as a distance metric to measure the similarity
between the patches~\citep{cvpr2003_Criminisi}. We denote the color
value of a pixel $x$ by $I_{x}=(R_{x},G_{x},B_{x})$. For an object
patch $\Psi_{p}$, the best patch $\Psi_{\tilde{q}}$ is chosen by
computing: 
\begin{equation}
\Psi_{\tilde{q}}=\arg\min_{q\in\Phi}d(\Psi_{p},\Psi_{q}),
\end{equation}
in which $d(\Psi_{q},\Psi_{p})$ is defined as follows: 
\begin{equation}
d(\Psi_{q},\Psi_{p})=\sum_{p\in\Psi_{p}\cap\Phi,q\in\Psi_{q}\cap\Phi}(I_{p}-I_{q})^{2}.
\end{equation}
Assume that the size of the original image is $M\times N$, and the
size of the patch is $P\times P$. The complexity of \textit{findBestPatch}
can be estimated as $O(MNP^2)$. Based on our profiling (will be shown
in Section 4), \emph{findBestPatch} compute kernel occupies the most
computations in the whole object removal algorithm.

Once the best matching patch is found, we copy the pixel values of
$\Psi_{\tilde{q}}$ into $\Psi_{p}$. The aforementioned search and
copy process is repeated until the whole target region $\Omega$ is
filled up. More details of the algorithm can be found in reference~\citep{cvpr2003_Criminisi}. 

\begin{figure*}
\begin{centering}
\includegraphics[width=0.8\textwidth]{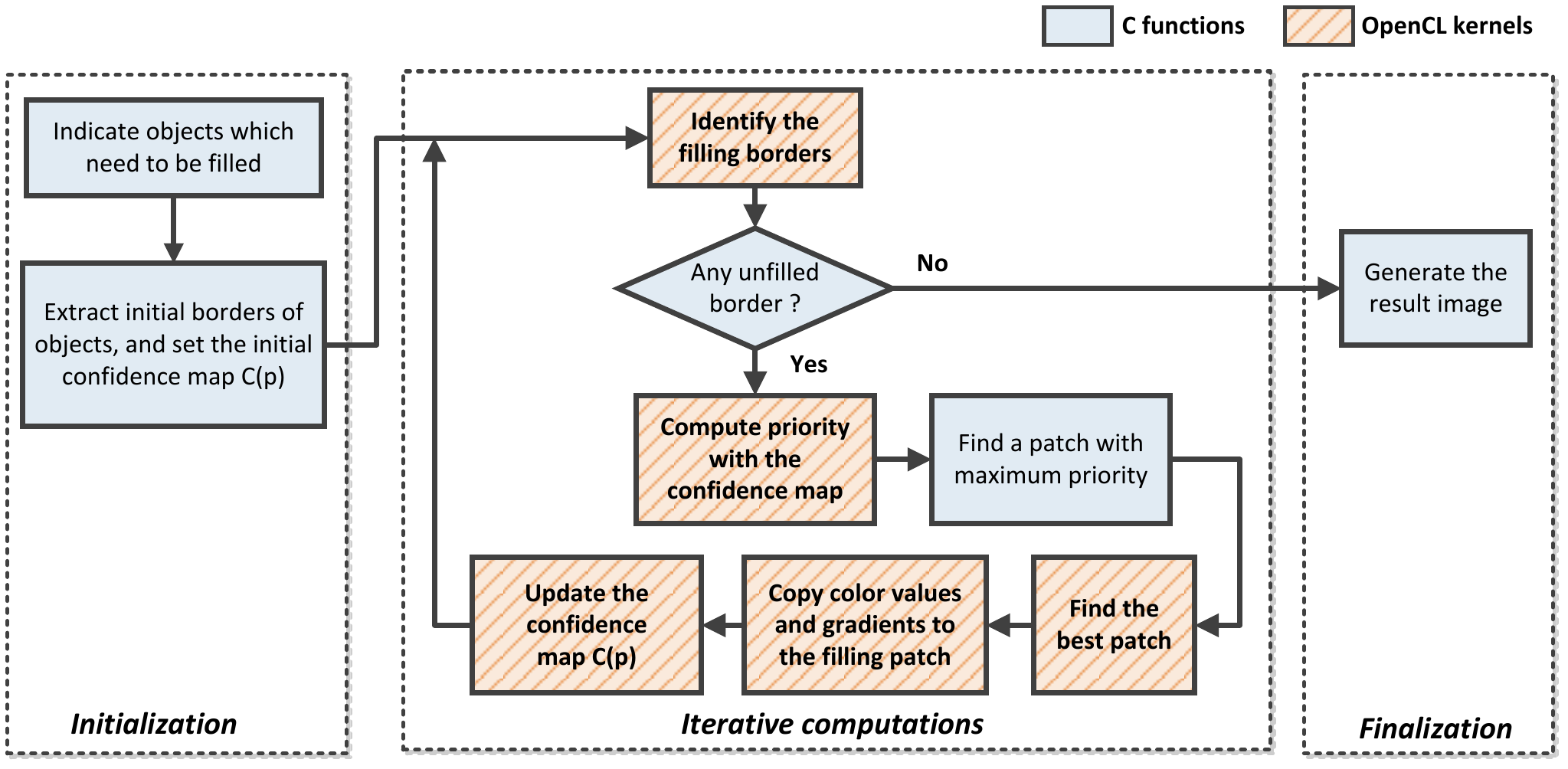}
\par\end{centering}

\caption{Algorithm workflow of the exemplar-based object removal algorithm.
Please note that one OpenCL block might be mapped to one or multiple
OpenCL kernels depending on the computation and memory data access
patterns.}

\label{fig:workflow}
\end{figure*}

\section{Algorithm Workflow Analysis and CPU-GPU Partitioning\label{sec:Workflow-Analysis}}

In this section, we analyze the workflow of the object removal algorithms
and describe the algorithm partitioning between the CPU and GPU to
fully utilize the resources of the mobile SoC chipset.

\subsection{Experimental Setup\label{sub:Experimental-Setup}}

The profiling and experiments are performed on a development platform
consisting of a Qualcomm Snapdragon 8974 chipset~\citep{Snapdragon},
which supports OpenCL Embedded Profile for both CPU and GPU. The details
of the experimental setup are listed in Table~\ref{tab:experimental_setup}.

\subsection{Algorithm Mapping\label{sub:Algorithm-Mapping}}

Fig. \ref{fig:workflow} shows a workflow diagram of the exemplar-based
inpainting algorithm for object removal. The algorithm can be partitioned
into three stages: initialization stage, iterative computation stage,
and the finalization stage. The blocks with the slashed lines are
core functions inside the iterative stage and represent most of the
computational workload. We can map the core functions into OpenCL
kernels to exploit the 2-dimensional pixel-level and block-level parallelisms
in the algorithms. The CPU handles the OpenCL context initialization,
memory objects management, and kernel launching. By analyzing the
algorithm, we partition the core functions into eight OpenCL kernels
based on the properties of the computations, as is shown in Table~\ref{table:breakdown}.
In each OpenCL kernel, the fact that no dependency exists among image
blocks allows us to naturally partition the tasks into work groups.
To represent color pixel values in RGBA (red green blue alpha) format,
we use efficient vector data structures such as \textit{cl\_uchar4}
to take advantage of built-in vector functions of OpenCL.

\begin{table}
\caption{Breakdown of execution time for OpenCL kernel functions running only
on CPU.}

\scriptsize

\begin{centering}
\begin{tabular}{c|c|c}
\hline 
\multirow{2}{*}{Kernel functions} & \multicolumn{1}{c|}{Exec } & \multirow{2}{*}{\%}\tabularnewline
 & time {[}s{]} & \tabularnewline
\hline 
\hline 
Convert RGB image to gray-scale image & 0.08 & 0.09\%\tabularnewline
\hline 
Update border of the area to be filled & 0.60 & 0.69\%\tabularnewline
\hline 
Mark source pixels to be used & 0.66 & 0.76\%\tabularnewline
\hline 
Update pixel priorities in the filling area & 0.45 & 0.52\%\tabularnewline
\hline 
Update pixel confidence in the filling area & 0.36 & 0.41\%\tabularnewline
\hline 
\textbf{Find the best matching patch} & \textbf{84.4} & \textbf{97.0\%}\tabularnewline
\hline 
Update RGB and grayscale image  & \multirow{2}{*}{0.46} & \multirow{2}{*}{0.53\%}\tabularnewline
of the filling patch &  & \tabularnewline
\hline 
\hline 
Total Time & 87.0 & 100\%\tabularnewline
\hline 
\end{tabular}
\par\end{centering}

\normalsize

\label{table:breakdown}
\end{table}

To better optimize the OpenCL-based implementation, we first measure
the timing performance of the OpenCL kernels. Table~\ref{table:breakdown}
shows a breakdown of processing time when running the program on a
single core of the CPU on our test device. The OpenCL kernel function
used to find the best matching patch with the current patch (denoted
as \textit{findBestPatch}) occupies most of the processing time (97\%),
so optimizing the \textit{findBestPatch} kernel becomes the key to
improving performance.

\section{Algorithm Optimizations and Implementation Trade-offs\label{sec:Algorithm-Optimizations}}

\subsection{OpenCL Implementation of \emph{findBestPatch} Kernel Function}

\begin{figure}
\begin{centering}
\includegraphics[width=1\columnwidth]{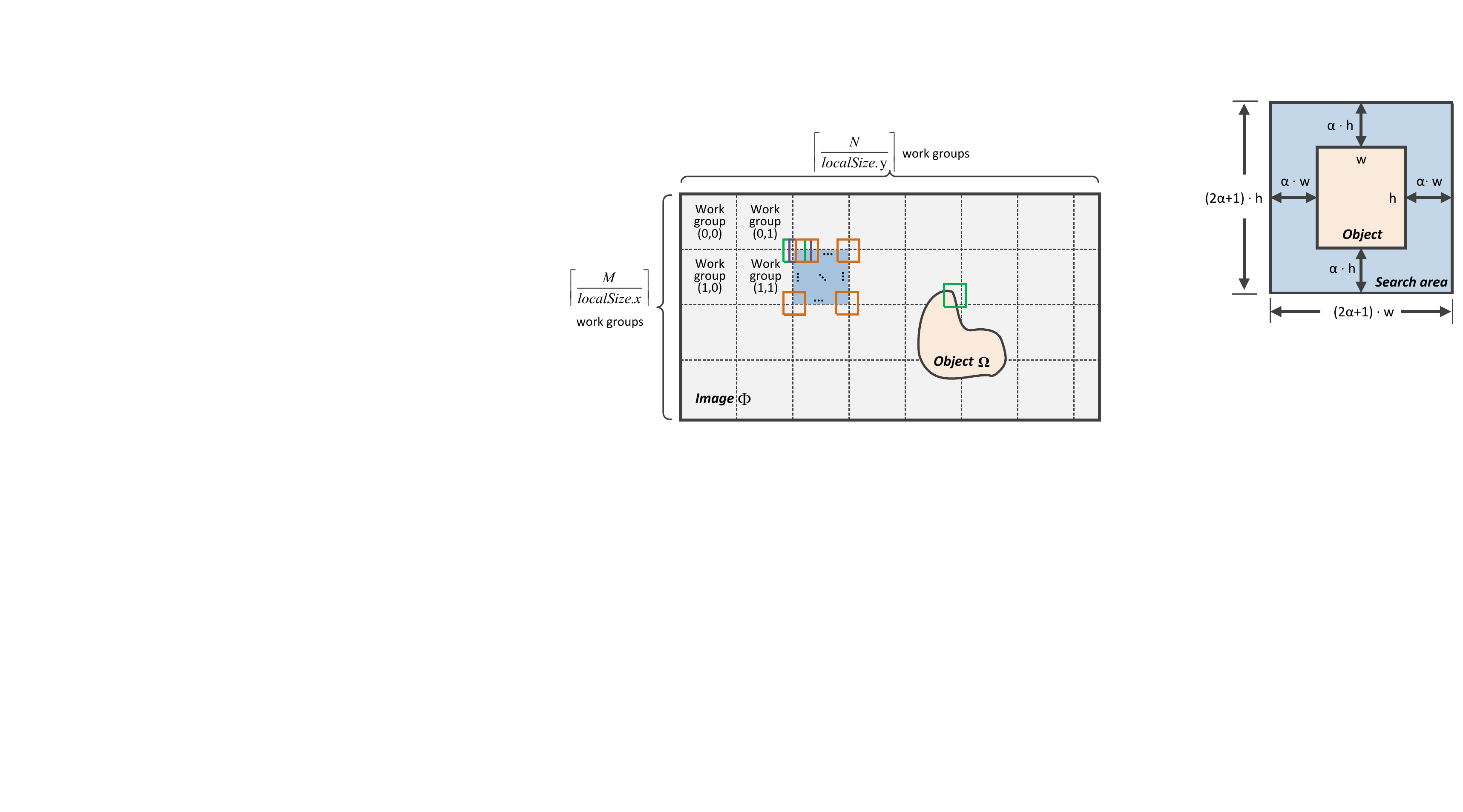}
\par\end{centering}

\caption{Algorithm mapping of \emph{findBestPatch} kernel function using OpenCL.
\label{fig:Algorithm-mapping}}
\end{figure}

\begin{algorithm}
\small

\caption{Compute SSD values between all candidate image patches and the image
patch to be filled using an OpenCL kernel function.\label{alg:findBestPatch-kernel-alg}}

\begin{enumerate}
\item \textbf{Input}: 

\begin{enumerate}
\item Position of object patch $\Psi_{p}$: ($px$, $py$);
\item Patch size $P$;
\end{enumerate}
\item \textbf{Output}: SSD array $ssdArray$;
\item \textbf{Begin OpenCL kernel:}
\item $\quad$Get global ID for the current work item: ($i$, $j$);
\item $\quad$float $sum=0.0$;
\item $\quad$int $src_{x}$, $src_{y}$; // source pixel position
\item $\quad$int $tgt_{x}$, $tgt_{y}$; // target pixel position
\item $\quad$int $winsize=P/2$;
\item $\quad$\textbf{for}(int $h=-winsize$; $h\leqslant winsize$; $h++$)
\item $\quad$$\quad$\textbf{for}(int $w=-winsize$; $w\leqslant winsize$;
$w++$)
\item $\quad$$\quad$$\quad$$src_{x}=i+w$; $src_{y}=j+h$;
\item $\quad$$\quad$$\quad$$tgt_{x}=px+w$; $tgt_{y}=py+h$;
\item $\quad$$\quad$$\quad$\textbf{if}(($src_{x}$, $src_{y}$) or ($tgt_{x}$,
$tgt_{y}$) is out of image) 
\item $\quad$$\quad$$\quad$$\quad$\textbf{continue};
\item $\quad$$\quad$$\quad$\textbf{end if}
\item $\quad$$\quad$$\quad$\textbf{if}(pixel ($tgt_{x}$, $tgt_{y}$)
is inside source region $\Phi$)
\item $\quad$$\quad$$\quad$$\quad$Read pixel ($tgt_{x}$, $tgt_{y}$)
data into $tgtData$;
\item $\quad$$\quad$$\quad$$\quad$Read pixel ($src_{x}$, $src_{y}$)
data into $srcData$;
\item $\quad$$\quad$$\quad$$\quad$$sum$$+=$$(tgtData-srcData)^{2}$;
\item $\quad$$\quad$$\quad$\textbf{end if}
\item $\quad$$\quad$\textbf{end for}
\item $\quad$\textbf{end for}
\item Store $sum$ into $ssdArray$;
\item \textbf{End OpenCL kernel}
\end{enumerate}
\normalsize
\end{algorithm}

\begin{table*}
\begin{centering}
\caption{Description of images in the test dataset. These images are selected
to represent different types of image scenes and different shapes
of objects to be removed. \label{tab:Description-of-images}}

\par\end{centering}

\centering{}%
\begin{tabular}{c|c|c|c|c}
\hline 
\textbf{Image} & \textbf{Image type} & \textbf{Image size} & \textbf{Object type} & \textbf{Object size}\tabularnewline
\hline 
\hline 
WalkingMan & Complex scene & $512\times384$ & Blob & $78\times126$\tabularnewline
\hline 
River & Complex scene, texture & $480\times639$ & Small blob & $59\times93$\tabularnewline
\hline 
DivingMan & Texture & $576\times384$ & Big/random shape & $155\times186$\tabularnewline
\hline 
Hill & Complex scene & $1024\times550$ & Long strip & $1024\times10$\tabularnewline
\hline 
\end{tabular}
\end{table*}

\begin{figure*}
\begin{centering}
\includegraphics[width=0.8\textwidth]{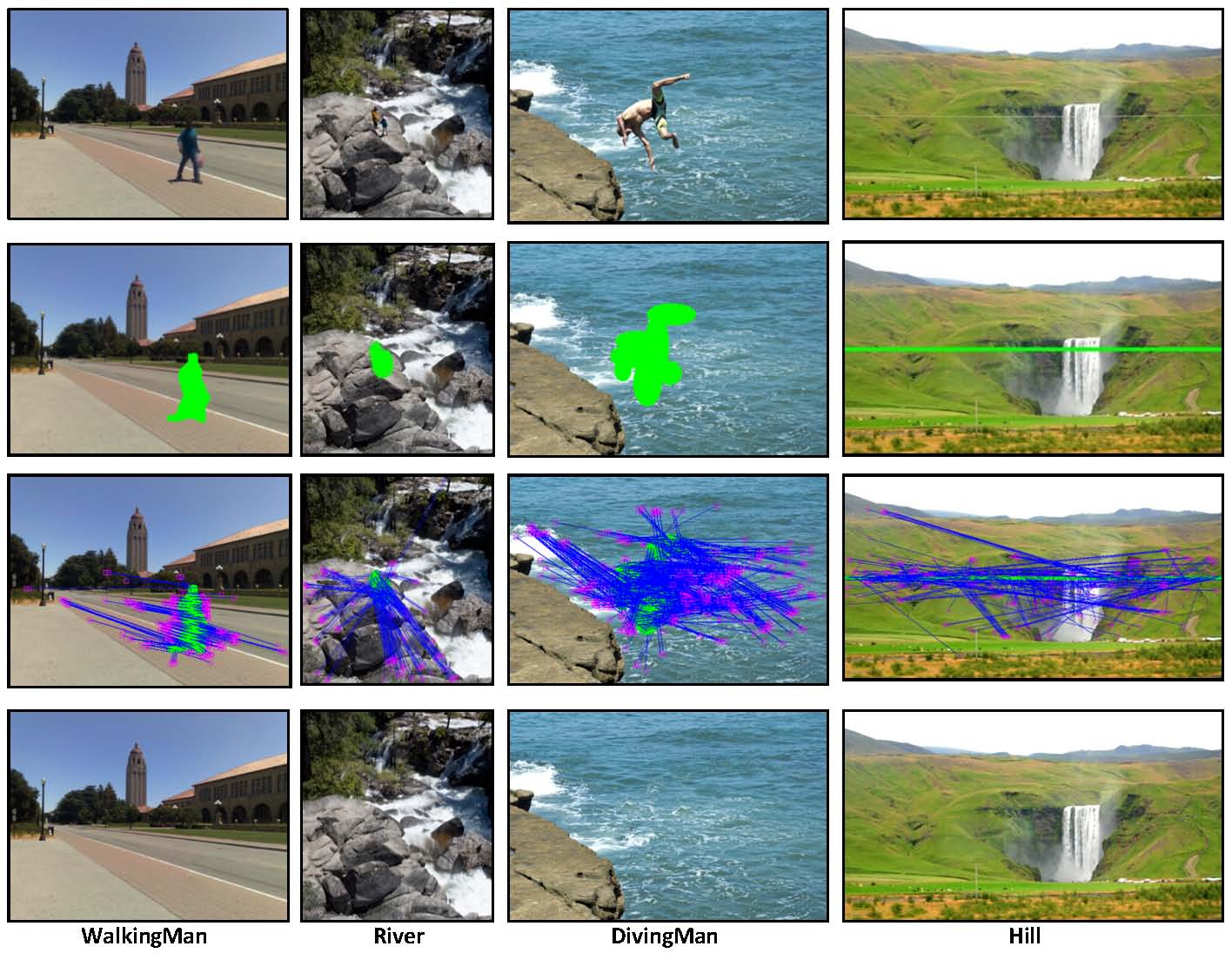}
\par\end{centering}

\caption{The best patch mapping found by a full search and the result images.
The 1st row: original images. The 2nd row: masks covering the unwanted
objects. The 3rd row: best patch mapping; the small squares indicate
the best patches found by the \emph{findBestPatch}() function. The
4th row: result images.\label{fig:dataset-images}}
\end{figure*}

The algorithm mapping of \emph{findBestPatch} kernel used for the
OpenCL implementation is shown in Fig.~\ref{fig:Algorithm-mapping}.
To perform a full search for the best patch $\Psi_{\tilde{q}}$ to
match the current filling patch $\Psi_{p}$ in the \textit{findBestPatch}
OpenCL kernel, we spawn $M\times N$ work items, with each computing
an SSD value between two $P\times P$ patches. We partition these
$M\times N$ work items into work groups according to the compute
capability of the GPU. The size of 2-dimensional work groups can be
expressed as

\begin{equation}
(\lceil M/localSize.x\rceil,\lceil N/localSize.y\rceil).
\end{equation}

In our implementation, each work group contains $8\times8$ work items
($localSize.x$$=$$8$, $localSize.y$$=\mbox{8}$). Therefore, each
work group of work items perform SSD computations for 64 patch candidates.
The parallel implementation of the SSD computation in the \textit{findBestPatch}
kernel function is detailed in Algorithm~\ref{alg:findBestPatch-kernel-alg}.

\subsection{Experimental Dataset}

To demonstrate the best patch matching behavior of the object removal
algorithm, we employ several different test images to cover different
scenarios, which are shown in the first row of Fig.~\ref{fig:dataset-images}.
The characteristics of these test images are summarized in Table~\ref{tab:Description-of-images}.
We chose images with different background scenes and textures, with
variable image sizes and object sizes, and with different object shapes,
so that by performing experiments on these images, we can better understand
the implementation trade-offs related to the performance improvement.

\subsection{Reducing Search Space }

\begin{figure}
\begin{centering}
\includegraphics[width=0.8\columnwidth]{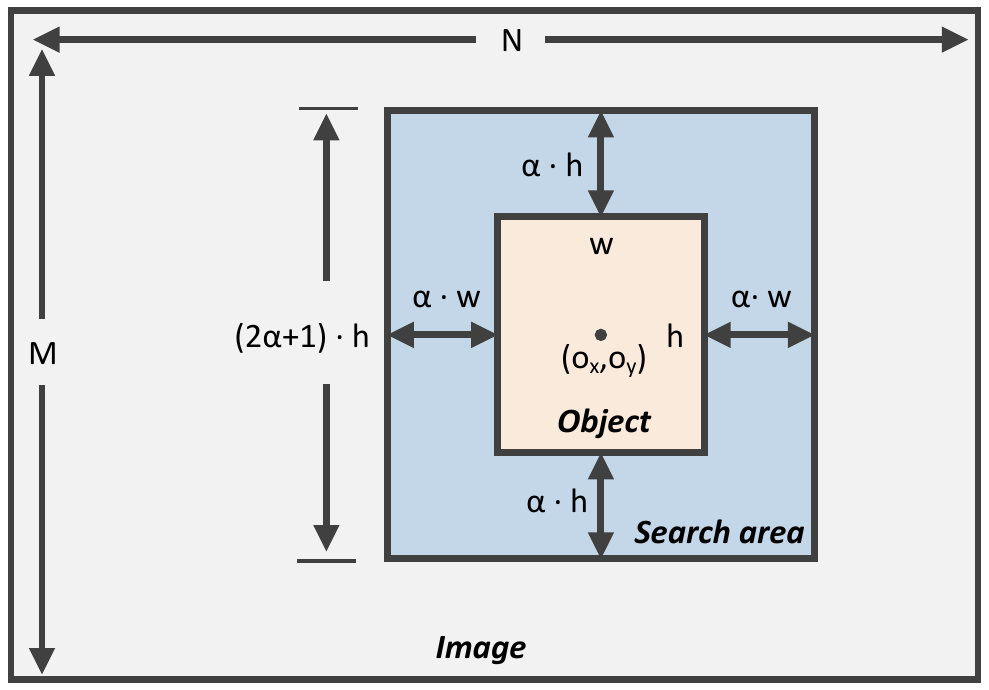}
\par\end{centering}

\caption{Diagram of reduced search area with search factor $\alpha$.\label{fig:reduced-search-area-diagram}}
\end{figure}

We have done experiments to verify the locations of the best patches
found by a full search across the whole image area. For the test images
shown in Fig.~\ref{fig:dataset-images}, most of the best patches
are found in a region surrounding the object area. The reason is that
adjacent areas usually have similar structures and textures in natural
images. To reduce the searching time, we can utilize this spatial
locality by limiting the search space. To better model this optimizing
strategy in a scalable manner, we define a search factor $\alpha$.
Assume the width and height of the object area are $w$ and $h$ respectively.
The new search area is formed by expanding the object area by $\alpha h$
to the up and down directions, and $\alpha w$ to the left and right
directions, as is shown in Fig.~\ref{fig:reduced-search-area-diagram}.
The search area factor $\alpha$ has a range of $0\leq\alpha<\max$
$(M/h,\, N/w)$. Assume the object region is centered at coordinate
$(o_{x,}o_{y}).$ Fig.~\ref{fig:reduced-search-area-diagram} shows
a typical case in which the whole search area is inside the image
area. For more general cases, the boundary of the new search area
becomes:

\begin{align}
B_{left} & =\max(0,o_{x}-(\frac{1}{2}+\alpha)w),\nonumber \\
B_{right} & =\min(N,o_{x}+(\frac{1}{2}+\alpha)w),\nonumber \\
B_{top} & =\max(0,o_{x}-(\frac{1}{2}+\alpha)h),\nonumber \\
B_{bottom} & =\min(M,o_{x}+(\frac{1}{2}+\alpha)h).
\end{align}

By defining the search factor $\alpha$ this way, we can easily adjust
the search area. Moreover, this method allows the search area to grow
along four directions with an equal chance, so as to increase the
possibility of finding a better patch. Since there are no useful pixels
in the object area for patch matching, only the candidate patches
not in the object region will be compared with the object patch. So
the actual size of the search area ($SA)$ can be expressed as:

\begin{eqnarray}
SA & = & (2\alpha+1)w\cdot(2\alpha+1)h-wh\nonumber \\
 & = & ((2\alpha+1)^{2}-1)wh.\label{eq:SA}
\end{eqnarray}
The complexity of \emph{findBestPatch} can be estimated by $O(((2\alpha+1)^{2}-1)whP^{2})$.
Thus, when we reduce the search area (reducing $\alpha$), the complexity
to search the best patch reduces significantly. 

\begin{figure*}
\begin{centering}
\includegraphics[width=1\textwidth]{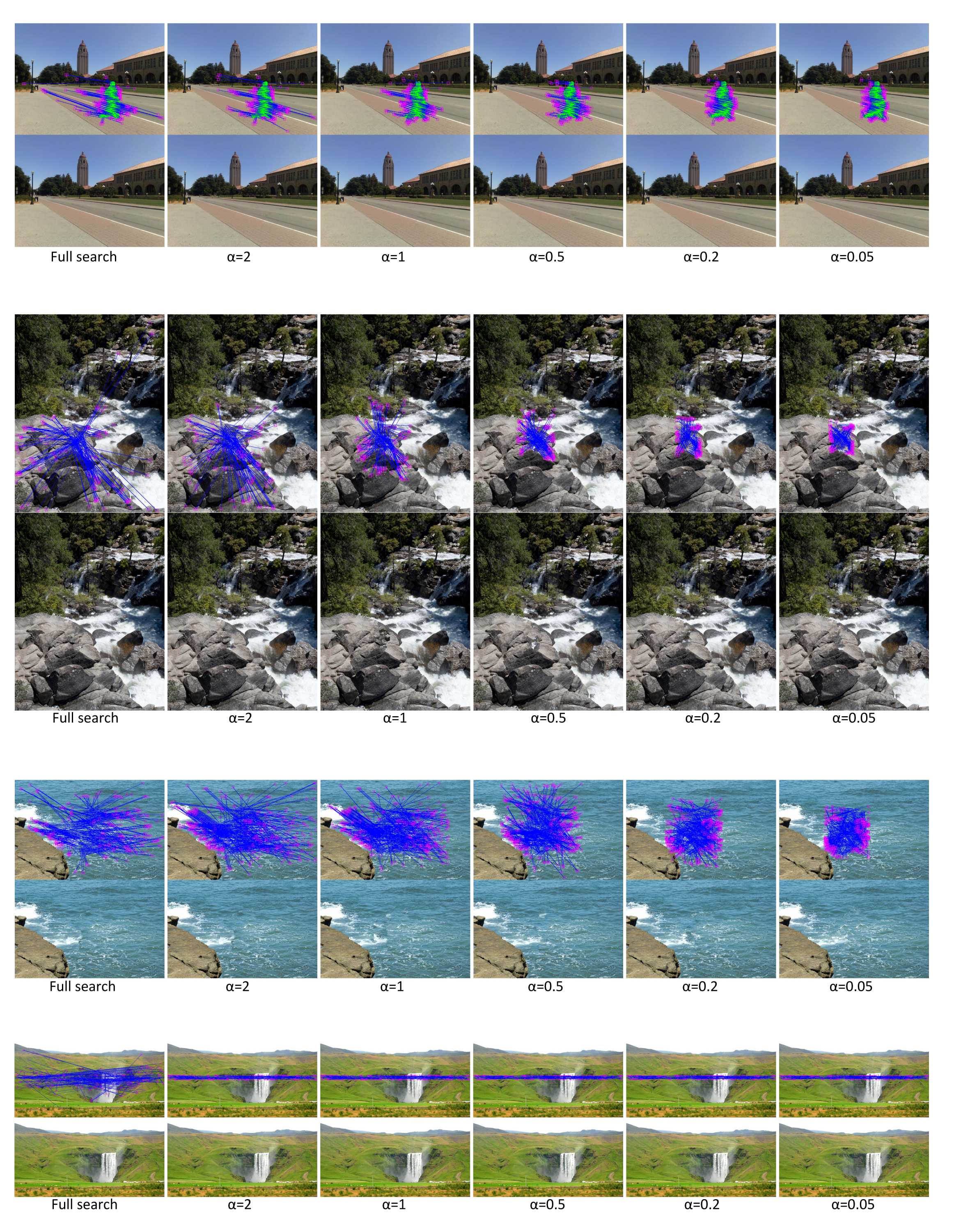}
\par\end{centering}

\caption{The effect of reducing the search area. The search area factor $\alpha$
is defined as in Fig.~\ref{fig:reduced-search-area-diagram}. \label{fig:effect-reducing-search-area}}
\end{figure*}

Fig.~\ref{fig:effect-reducing-search-area} demonstrates the effect
of reducing the search factor $\alpha$. After reducing the search
factor, the best matching patches are limited to a small region around
the object region. Even when the search factor is reduced significantly
to only $\alpha=0.05$, we still get visually plausible results. Due
to its importance, choosing a proper search factor $\alpha$ is critical
for practical applications to achieve good performance and efficiency.
Based on our experiments, for a regular object region, choosing parameter
$\alpha$ in the range of $0.05\leq\alpha<0.5$ normally leads to
good implementation trade-offs. If the algorithm is to be applied
to a certain type of images, trainings on datasets can be performed
to determine a proper search factor $\alpha$.

In addition to time reduction, reducing the search area can also reduce
the possible false matching. As a comparison metric, SSD can roughly
represent the similarity of two patches, but it cannot accurately
reflect the structural and color information embedded in the patches.
Therefore, the patches with the highest distance scores (SSD in this
algorithm) may not be the best patches to fill in the hole and to
generate visually plausible results due to the limitation of SSD metric,
especially for complex scenes with different color information and
structures. The algorithm sometimes can lead to false matching, in
which the reported ``best'' patch with a high correlation score
may have very distinctive textural or color information compared to
the object patch. Under such circumstances, the artificial effects
introduced by the false matching will degrade the quality of the result
images significantly. Fortunately, spatial locality can be observed
in most of the natural images, therefore, the visually plausible matching
patches (in terms of color and texture similarity) tend to reside
in the surrounding area of the candidate patch with high chances.
By reducing the search area to a certain degree, we can reduce the
possibility of false matching and therefore generate visually plausible
result images. 

\begin{figure}
\begin{centering}
\includegraphics[width=1\columnwidth]{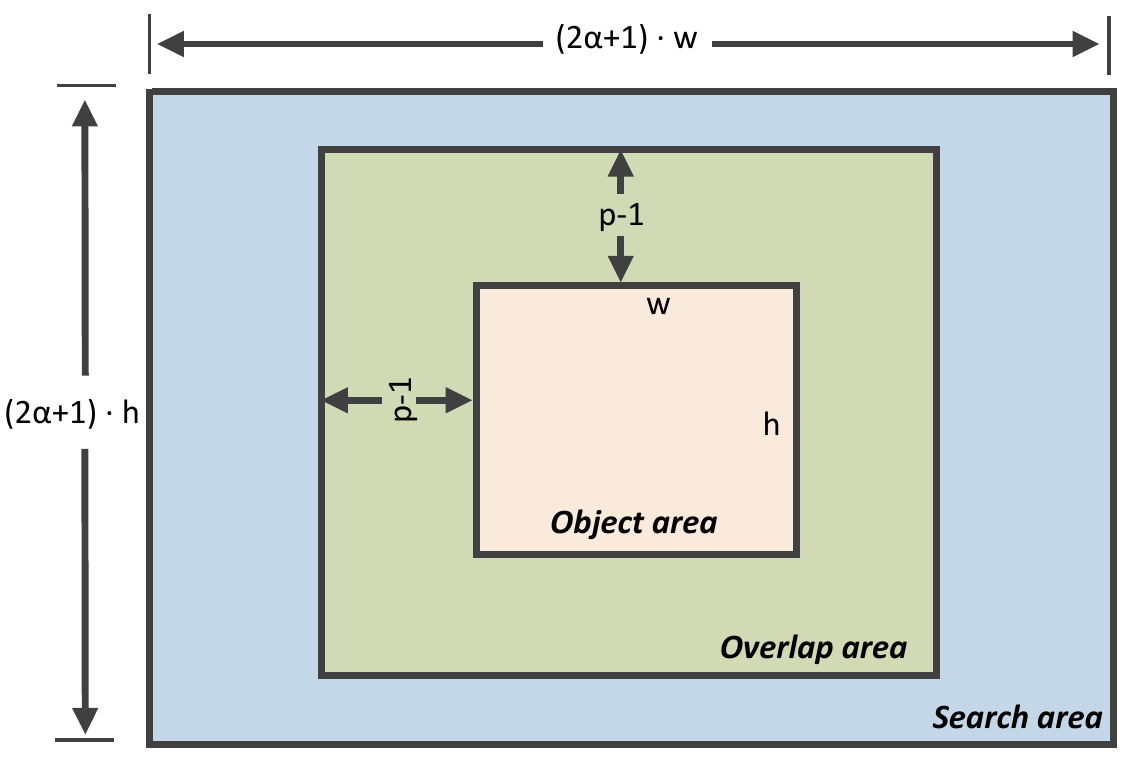}
\par\end{centering}

\caption{The relationship among search area, object area and the overlap area
for the best patch searching. \label{fig:increase-patch-size-diagram}}
\end{figure}

\begin{figure}
\begin{centering}
\includegraphics[width=1\columnwidth]{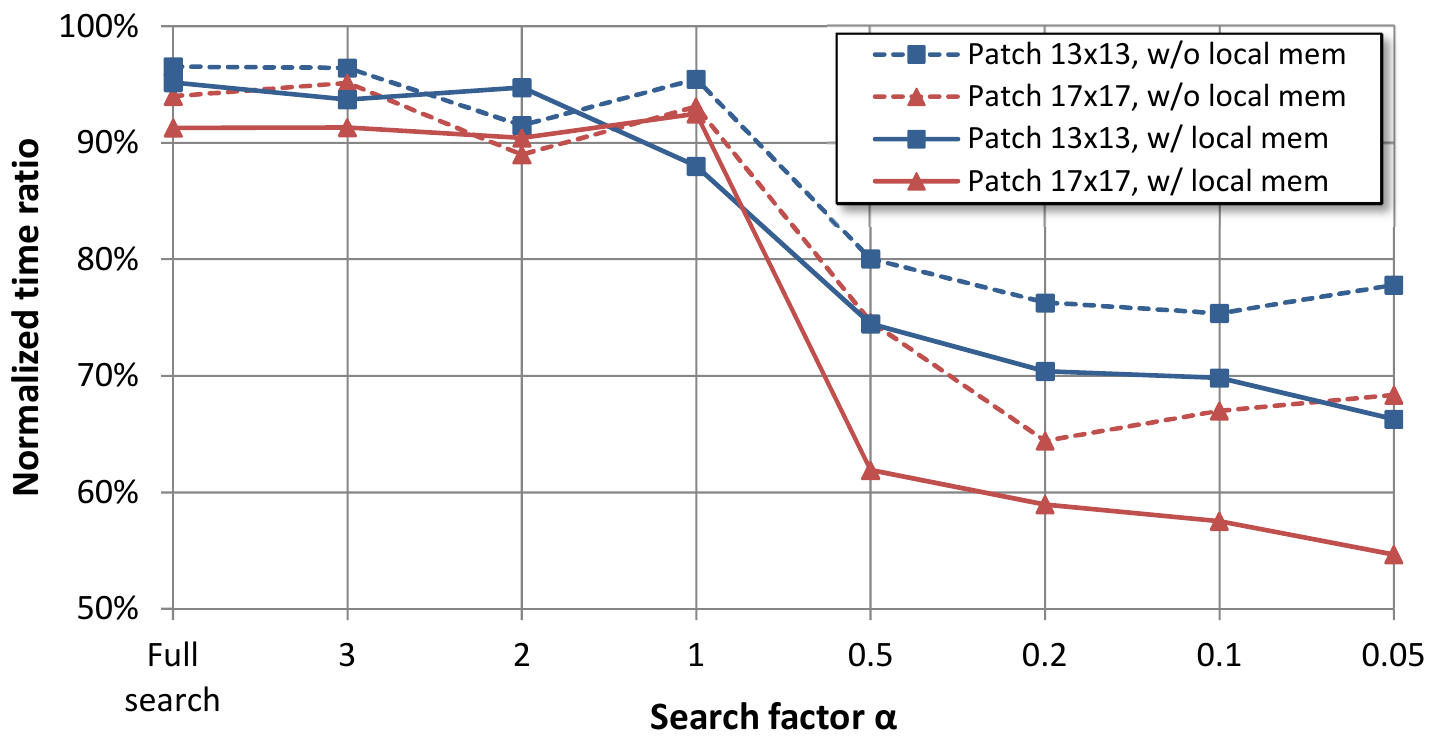}
\par\end{centering}

\caption{Impact of increased patch size $13\times13$ and $17\times17$. ``WalkingMan''
test image. The processing time for $13\times13$ and $17\times17$
patches is normalized by the time of the $9\times9$ patch. (The $9\times9$
patch size is suggested by the original algorithm proposed by Criminisi
\emph{et al.}~\citep{cvpr2003_Criminisi}). \label{fig:increase-patch-size-curve}}
\end{figure}

\begin{figure*}
\begin{centering}
\includegraphics[width=0.9\textwidth]{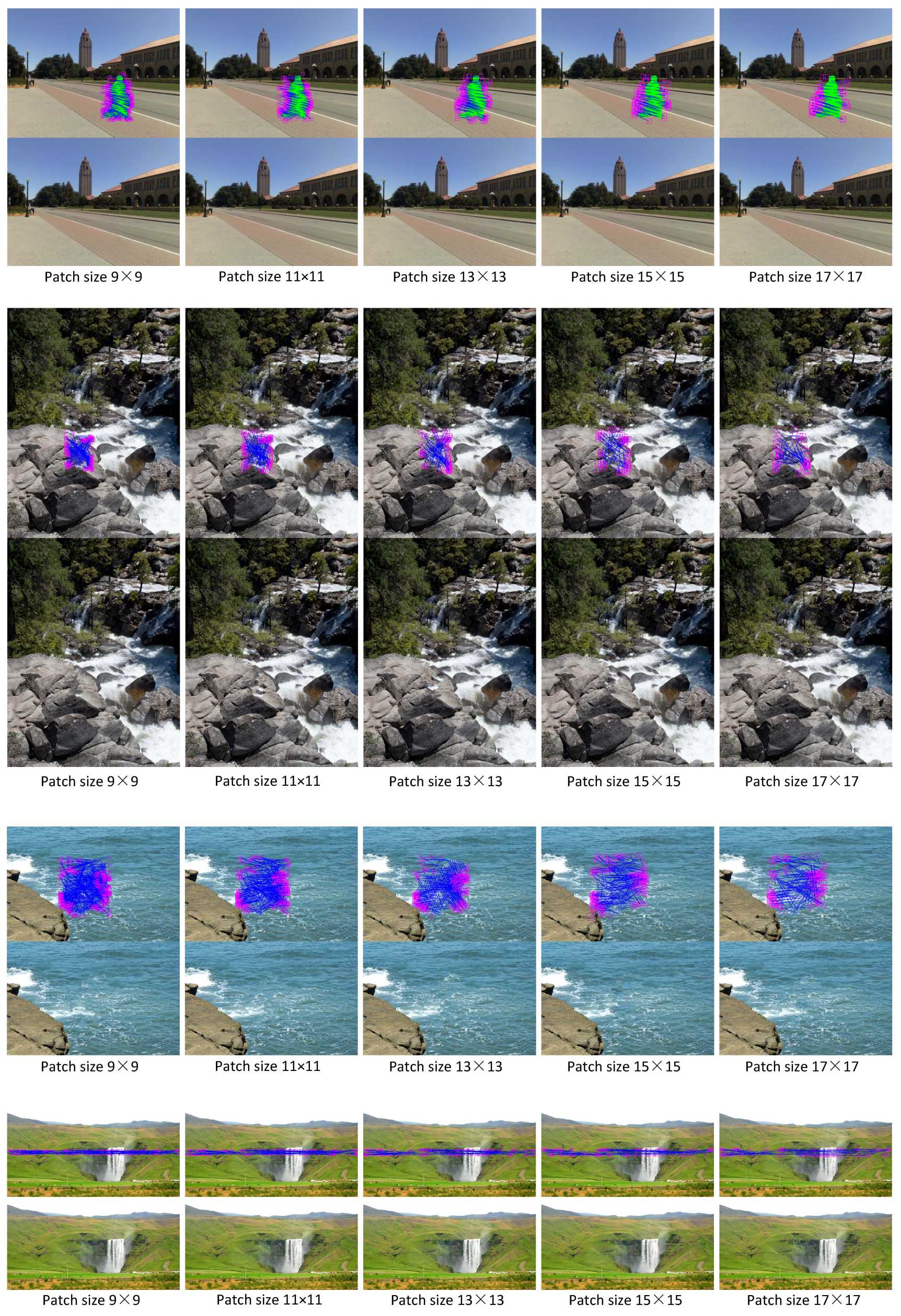}
\par\end{centering}

\caption{The experimental results for increased patch size. ($\alpha=0.05$
in this experiment.)\label{fig:effect-increase-patch-size}}
\end{figure*}

\subsection{Optimizing Patch Size}

The object removal algorithm is an iterative algorithm, in which one
object patch is processed in each iteration. We need roughly $wh/P^{2}$
iterations to finish the whole process. That said, if we increase
patch size $P$, fewer iterations are needed to fill the object area
which may lead to shorter processing time. Meanwhile, the complexity
of the SSD computation ($O(P^{2})$) becomes higher for the patch
matching, which tends to increase the processing time. Therefore,
it is not straightforward to determine the impact of increasing patch
size $P$ to the overall complexity and performance. 

We use Fig.~\ref{fig:increase-patch-size-diagram} to help us analyze
the overall computation complexity. First of all, we assume the search
factor $\alpha$ is defined as in the previous section. We also define
the search area as $SA$ as in (\ref{eq:SA}). 

Secondly, because the patch size is $P$, any candidate patch within
$(P-1)$ pixels range surrounding the object area $\Omega$ will partially
overlap with the object area. We define this overlapping area as $OA$,
whose area can be calculated as:

\begin{equation}
OA=(2(P-1)+w)\cdot(2(P-1)+h)-wh.
\end{equation}

If the candidate patch lies outside the overlapping area%
\footnote{In this case, the candidate patch is in the area of $(SA-OA)$.%
}, the complexity of the SSD computation can be estimated as $O(P^{2})$.
When the candidate patch and the object area overlaps%
\footnote{In this case, the candidate patch is in the area of $OA$.%
}, we only use the pixels $I_{q}$ in the intersection of the candidate
patch $\Psi_{q}$ and source image $\Phi$ ($I_{q}\in\Psi_{q}\cap\Phi$)
to perform the computation. We can estimate the computation complexity
as $O(kP^{2})$, in which $k$ is a constant value, representing the
average ratio of the overlapping area. For a typical rectangle search
area and object area, $k$ can be approximately estimated as 0.5.
Therefore, the overall computation complexity can be calculated as
in (\ref{eq:complexity_overall}).\\
\begin{figure*}[!t]
\footnotesize

\begin{eqnarray}
Complexity_{overall} & \approx & O(wh/P^{2})\cdot((2\alpha+1)^{2}-1)wh\cdot[Prob(\Psi_{q}\in(SA-OA))\cdot O(P{}^{2})+Prob(\Psi_{q}\in OA)\cdot O(kP^{2})]\nonumber \\
 & = & O[((2\alpha+1)^{2}-1)w^{2}h^{2}\cdot(\frac{SA-OA}{SA}+k\frac{OA}{SA})]\nonumber \\
 & = & O[((2\alpha+1)^{2}-1)w^{2}h^{2}\cdot(\frac{(2\alpha+1)^{2}wh-(2(P-1)+w)(2(P-1)+h)}{((2\alpha+1)^{2}-1)wh}\nonumber \\
 &  & +k\frac{(2(P-1)+w)(2(P-1)+h)-wh}{((2\alpha+1)^{2}-1)wh})]\nonumber \\
 & = & O[wh\cdot((2\alpha+1)^{2}wh-(2(P-1)+w)(2(P-1)+h)+k((2(P-1)+w)(2(P-1)+h)-wh))]\nonumber \\
 & = & O[wh\cdot(((2\alpha+1)^{2}-k)wh-(1-k)(2(P-1)+w)(2(P-1)+h))]\nonumber \\
 & = & O[w^{2}h^{2}\cdot((2\alpha+1)^{2}-(1-k)(2\frac{(P-1)}{w}+1)(2\frac{(P-1)}{h}+1)-k)].\label{eq:complexity_overall}
\end{eqnarray}

\hrulefill
\vspace*{4pt}
\end{figure*}
\normalsize

Experimental results show that the processing time can be reduced
by increasing patch size while reducing the search area. From Fig.~\ref{fig:increase-patch-size-diagram},
an intuitive explanation is that when we reduce the search area, more
candidate patches overlap with the object region. In this scenario,
the bigger the patches are, the more overlap there will be. As a result,
the amount of required computations becomes smaller. Thus, as we increase
the patch size and reduce the search area, the processing time decreases.
Equation~\eqref{eq:complexity_overall} shows that for bigger search
factor $\alpha$, the term $(2\alpha+1)^{2}$dominates the complexity.
In this case, the complexity change caused by the increased patch
size is negligible. However, when $\alpha$ becomes smaller, the search
area decreases. When the search area $SA$ becomes comparable to or
even smaller than the object area $\Omega$, the term $(2\alpha+1)^{2}$
becomes less dominant. Therefore, for smaller search factor $\alpha$,
increasing the patch size $P$ can reduce the computation complexity.
Experimental results shown in Fig.~\ref{fig:increase-patch-size-curve}
verify the above analysis. In Fig.~\ref{fig:increase-patch-size-curve},
the processing time for patch sizes of $13\times13$ and $17\times17$
is normalized by the processing time of the $9\times9$ patch to show
the performance speedup achieved by increasing the patch size. Experimental
results show that, for bigger search areas ($\alpha\geq1$), patch
size does not affect the performance. However, as the search area
keeps decreasing ($\alpha<1$), bigger patch size leads to more significant
time reduction. The experimental results in Fig.~\ref{fig:increase-patch-size-curve}
also indicate that increasing the patch size works for the experiments
with (or without) the memory optimization discussed in Section~\ref{sub:Memory-Optimization}.
In addition, Fig.~\ref{fig:effect-increase-patch-size} shows that
by increasing the patch size, we can generate visually plausible result
images without degrading the image quality.

\subsection{Memory Optimization\label{sub:Memory-Optimization}}

\begin{figure}
\begin{centering}
\includegraphics[width=0.9\columnwidth]{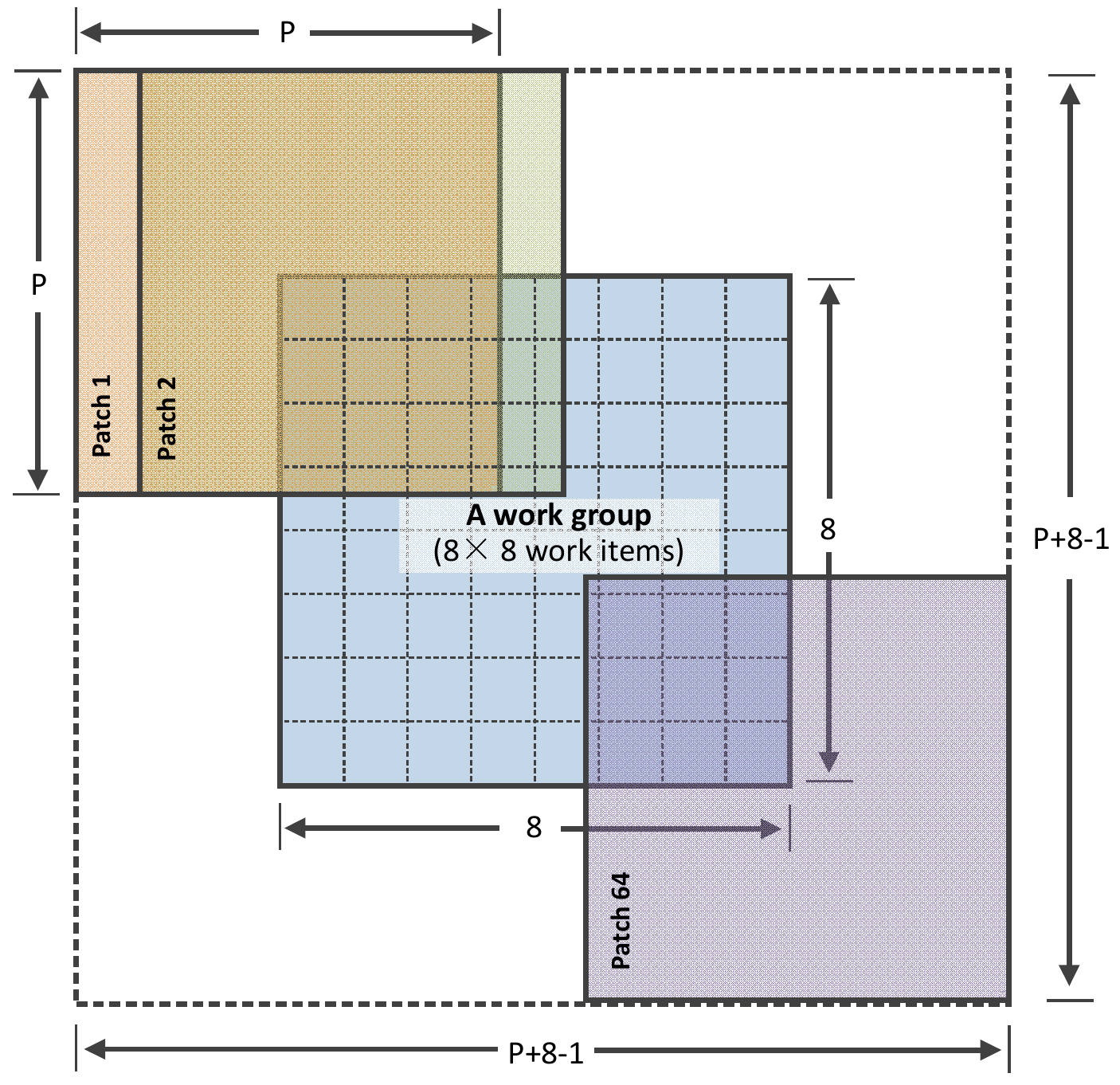}
\par\end{centering}

\caption{A detailed diagram showing how $8\times8$ work items in a work group
compute SSD values in parallel. Most of the pixels in the dashed box
are accessed multiple times by parallel work items.\label{fig:why-local-mem}}
\end{figure}

\begin{table}
\caption{Local memory usage for ``WalkingMan'' image. \label{tab:Local-memory-usage}}

\centering{}%
\begin{tabular}{c|c|c}
\hline 
\textbf{Patch size} & \textbf{Data} & \textbf{Local memory}\tabularnewline
\hline 
 & Source data & 1024 bytes\tabularnewline
\cline{2-3} 
$\mathbf{9\times9}$ & Patch data & 324 bytes\tabularnewline
\cline{2-3} 
$(p+8-1=16)$ & Patch pixel label & 324 bytes\tabularnewline
\cline{2-3} 
 & \textbf{Total} & \textbf{1672 bytes}\tabularnewline
\hline 
 & Source data & 1600 bytes\tabularnewline
\cline{2-3} 
\textbf{$\mathbf{13\times13}$} & Patch data & 676 bytes\tabularnewline
\cline{2-3} 
$(p+8-1=20)$ & Patch pixel label & 676 bytes\tabularnewline
\cline{2-3} 
 & \textbf{Total} & \textbf{2952 bytes}\tabularnewline
\hline 
 & Source data & 2304 bytes\tabularnewline
\cline{2-3} 
\textbf{$\mathbf{17\times17}$} & Patch data & 1156 bytes\tabularnewline
\cline{2-3} 
$(p+8-1=24)$ & Patch pixel label & 1156 bytes\tabularnewline
\cline{2-3} 
 & \textbf{Total} & \textbf{4616 bytes}\tabularnewline
\hline 
\end{tabular}
\end{table}

Similar to desktop GPUs, mobile GPUs also suffer from long latency
of the off-chip global memory. The local memory on the mobile GPU
provides fast memory accesses and can be shared by work items in the
same work group. As mentioned before, a work group contains $8\times8$
work items, each of which computes an SSD value between an object
patch and a candidate patch. As shown in Fig.~\ref{fig:why-local-mem},
the $n$-th work item works on the $n$-th patch. Most of the pixels
in adjacent candidate patches overlap. For instance, patch 1 and patch
2 share most of the image pixels and only one column of pixels in
each patch are different. Similarly, all adjacent candidate patches
processed by $8\times8$ work items have many overlapped pixels, each
of which is accessed multiple times by several different work items.
These unnecessary memory accesses to the global memory can lead to
long latency and increase the processing time. We can also tell from
Fig.~\ref{fig:why-local-mem} that for a $P\times P$ patch, $(P+8-1)\times(P+8-1)$
pixels are actually shared among work items. Thus, we can load these
pixels into the on-chip local memory to allow data sharing and avoid
unnecessary accesses to the global memory. In our OpenCL implementation,
$(P+8-1)^{2}\cdot sizeof$ $(cl\_uchar4)$ source image data, $P^{2}\cdot sizeof$
$(cl\_uchar4)$ patch image data and $P^{2}\cdot sizeof$ $(cl\_int)$
patch pixel label data can be loaded into the local memory. 

\begin{algorithm}[b]
\small

\caption{Parallel data loading from global memory to local memory in \emph{findBestPatch}
kernel function.\label{alg:Parallel-data-loading}}
\label{alg:local_mem}
\begin{enumerate}
\item Get global ID of the current work item: ($gid.x$, $gid.y$);
\item Get local ID of the current work item: ($lid.x$, $lid.y$);
\item $local\_id=lid.y*lsize.x+lid.x$;
\item Get local work group size: ($lsize.x$, $lsize.y$);
\item $group\_size=lsize.x*lsize.y$;
\item Calculate local memory size: $local\_mem\_size$;
\item \textbf{while}($local\_index<local\_mem\_size$)
\item $\quad$Calculate global memory address;
\item $\quad$\textbf{if}($local\_index<P*P$)
\item $\quad$$\quad$Calculate patch data address;
\item $\quad$$\quad$Load $patchPixelLabel$ into local memory;
\item $\quad$$\quad$Load $patchData$ into local memory;
\item $\quad$\textbf{end if}
\item $\quad$Load $srcData$ into local memory;
\item $\quad$$local\_index+=group\_size$;
\item \textbf{end while}
\item barrier(CLK\_LOCAL\_MEM\_FENCE);
\item Start SSD computation from here. (Similar to Algorithm 1, just read
needed data from local memory.)
\end{enumerate}
\normalsize
\end{algorithm}

\begin{figure}
\begin{centering}
\includegraphics[width=1\columnwidth]{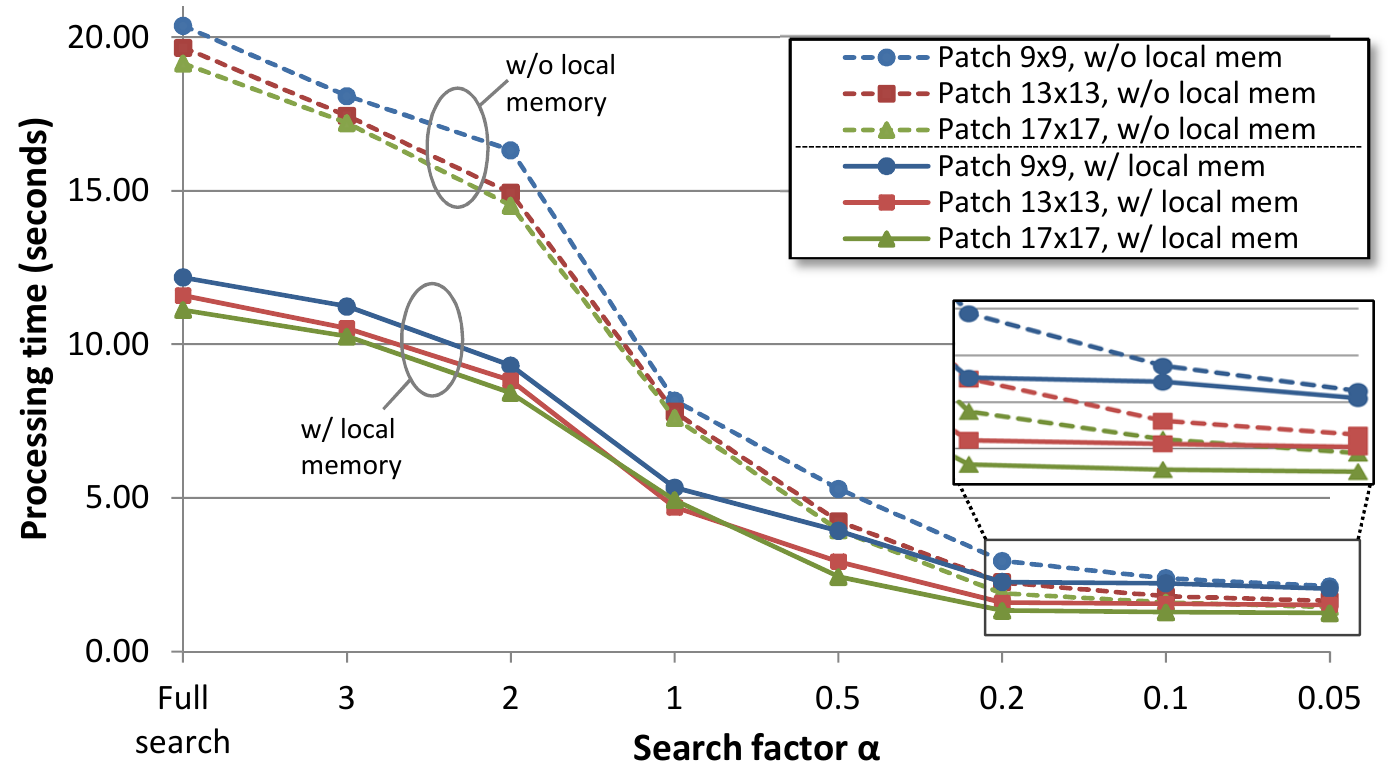}
\par\end{centering}

\caption{Performance comparison between using the local memory and without
using the local memory. The results for the ``WalkingMan'' are shown.
The experiments performed on other test images generate similar results.
\label{fig:local_mem_results}}
\end{figure}

\begin{table*}
\small

\caption{Total processing time for ``WalkingMan'' image, with OpenCL kernels
running on the GPU.}

\begin{centering}
\begin{tabular}{c|c|c|c|c|c|c|c|c|c|c}
\hline 
 &  & \multicolumn{3}{c|}{CPU-only} & \multicolumn{6}{c}{Heterogeneous CPU+GPU}\tabularnewline
\cline{6-11} 
Search factor $\alpha$ & Search area & \multicolumn{3}{c|}{} & \multicolumn{3}{c|}{w/o local memory ($s$)} & \multicolumn{3}{c}{w/ local memory ($s$)}\tabularnewline
\cline{3-11} 
 &  & \multicolumn{3}{c|}{Patch size} & \multicolumn{3}{c|}{Patch size} & \multicolumn{3}{c}{Patch size}\tabularnewline
\cline{3-11} 
 &  & $9\times9$ & $13\times13$ & $17\times17$ & $9\times9$ & $13\times13$ & $17\times17$ & $9\times9$ & $13\times13$ & $17\times17$\tabularnewline
\hline 
\hline 
full search & $512\times384$ & 23.26 & 22.48 & 23.68 & 20.37 & 19.66 & 19.15 & 12.18 & 11.59 & 11.12\tabularnewline
\hline 
2 & $382\times384$ & 18.08 & 16.97 & 18.30 & 16.32 & 14.93 & 14.52 & 9.31 & 8.82 & 8.42\tabularnewline
\hline 
1 & $234\times311$ & 13.37 & 9.91 & 9.74 & 8.17 & 7.80 & 7.61 & 5.34 & 4.69 & 4.94\tabularnewline
\hline 
0.5 & $156\times248$ & 11.27 & 9.80 & 8.20 & 5.29 & 4.24 & 3.95 & 3.93 & 2.93 & 2.43\tabularnewline
\hline 
0.2 & $109\times176$ & 7.13 & 5.20 & 3.79 & 2.95 & 2.25 & 1.90 & 2.26 & 1.59 & 1.33\tabularnewline
\hline 
0.05 & $86\times139$ & 6.02 & 4.73 & 3.85 & 2.12 & 1.65 & 1.45 & 2.04 & 1.52 & 1.25\tabularnewline
\hline 
\end{tabular}
\par\end{centering}

\normalsize

\label{table:experimental_results_walkingman}
\end{table*}

\begin{table*}
\small

\caption{Total processing time for ``River'' image. With OpenCL kernels running
on the GPU.}

\begin{centering}
\begin{tabular}{c|c|c|c|c|c|c|c|c|c|c}
\hline 
 &  & \multicolumn{3}{c|}{CPU-only} & \multicolumn{6}{c}{Heterogeneous CPU+GPU}\tabularnewline
\cline{6-11} 
Search factor $\alpha$ & Search area & \multicolumn{1}{c}{} & \multicolumn{1}{c}{} &  & \multicolumn{3}{c|}{Time w/o local memory ($s$)} & \multicolumn{3}{c}{Time w/ local memory ($s$)}\tabularnewline
\cline{3-11} 
 &  & \multicolumn{3}{c|}{Patch size} & \multicolumn{3}{c|}{Patch size} & \multicolumn{3}{c}{Patch size}\tabularnewline
\cline{3-11} 
 &  & $9\times9$ & $13\times13$ & $17\times17$ & $9\times9$ & $13\times13$ & $17\times17$ & $9\times9$ & $13\times13$ & $17\times17$\tabularnewline
\hline 
\hline 
full search & $480\times639$ & 25.94 & 25.01 & 27.41 & 18.28 & 16.19 & 16.01 & 14.77 & 13.80 & 13.35\tabularnewline
\hline 
2 & $295\times465$ & 12.23 & 12.15 & 12.68 & 8.52 & 7.93 & 7.95 & 6.71 & 6.94 & 6.59\tabularnewline
\hline 
1 & $177\times279$ & 7.76 & 6.98 & 5.34 & 5.09 & 3.13 & 3.11 & 4.15 & 2.74 & 2.60\tabularnewline
\hline 
0.5 & $118\times186$ & 4.80 & 3.91 & 3.82 & 2.81 & 2.06 & 1.42 & 2.34 & 1.87 & 1.5\tabularnewline
\hline 
0.2 & $83\times130$ & 3.21 & 1.66 & 1.57 & 1.93 & 1.18 & 1.05 & 1.84 & 1.34 & 1.06\tabularnewline
\hline 
0.05 & $65\times102$ & 2.29 & 1.72 & 1.91 & 1.39 & 1.15 & 1.01 & 1.66 & 1.16 & 1.00\tabularnewline
\hline 
\end{tabular}
\par\end{centering}

\normalsize

\label{table:experimental_results-river}
\end{table*}

\begin{table*}
\small

\caption{Total processing time for ``Dive'' image. With OpenCL kernels running
on the GPU.}

\begin{centering}
\begin{tabular}{c|c|c|c|c|c|c|c|c|c|c}
\hline 
 &  & \multicolumn{3}{c|}{CPU-only} & \multicolumn{6}{c}{Heterogeneous CPU+GPU}\tabularnewline
\cline{6-11} 
Search factor $\alpha$ & Search area & \multicolumn{1}{c}{} & \multicolumn{1}{c}{} &  & \multicolumn{3}{c|}{Time w/o local memory ($s$)} & \multicolumn{3}{c}{Time w/ local memory ($s$)}\tabularnewline
\cline{3-11} 
 &  & \multicolumn{3}{c|}{Patch size} & \multicolumn{3}{c|}{Patch size} & \multicolumn{3}{c}{Patch size}\tabularnewline
\cline{3-11} 
 &  & $9\times9$ & $13\times13$ & $17\times17$ & $9\times9$ & $13\times13$ & $17\times17$ & $9\times9$ & $13\times13$ & $17\times17$\tabularnewline
\hline 
\hline 
full search & $576\times384$ & 64.48 & 62.86 & 61.47 & 44.83 & 40.45 & 36.65 & 44.62 & 34.97 & 31.08\tabularnewline
\hline 
2 & $576\times384$ & 65.11 & 64.83 & 63.55 & 44.73 & 40.29 & 36.61 & 44.67 & 35.00 & 31.11\tabularnewline
\hline 
1 & $465\times384$ & 52.17 & 54.31 & 54.99 & 34.22 & 33.26 & 19.33 & 28.62 & 28.65 & 27.81\tabularnewline
\hline 
0.5 & $310\times369$ & 54.36 & 37.07 & 36.82 & 21.78 & 22.73 & 8.24 & 19.98 & 23.04 & 16.68\tabularnewline
\hline 
0.2 & $217\times260$ & 40.92 & 29.64 & 27.69 & 15.14 & 9.45 & 5.66 & 14.01 & 10.40 & 7.40\tabularnewline
\hline 
0.05 & $171\times205$ & 35.44 & 22.22 & 20.171 & 13.74 & 7.90 & 5.40 & 14.21 & 7.84 & 5.31\tabularnewline
\hline 
\end{tabular}
\par\end{centering}

\normalsize

\label{table:experimental_results-dive}
\end{table*}

\begin{table*}
\small

\caption{Total processing time for ``Hill'' image. With OpenCL kernels running
on the GPU.}

\begin{centering}
\begin{tabular}{c|c|c|c|c|c|c|c|c|c|c}
\hline 
 &  & \multicolumn{3}{c|}{CPU-only} & \multicolumn{6}{c}{Heterogeneous CPU+GPU}\tabularnewline
\cline{6-11} 
Search factor $\alpha$ & Search area & \multicolumn{1}{c}{} & \multicolumn{1}{c}{} &  & \multicolumn{3}{c|}{Time w/o local memory ($s$)} & \multicolumn{3}{c}{Time w/ local memory ($s$)}\tabularnewline
\cline{3-11} 
 &  & \multicolumn{3}{c|}{Patch size} & \multicolumn{3}{c|}{Patch size} & \multicolumn{3}{c}{Patch size}\tabularnewline
\cline{3-11} 
 &  & $9\times9$ & $13\times13$ & $17\times17$ & $9\times9$ & $13\times13$ & $17\times17$ & $9\times9$ & $13\times13$ & $17\times17$\tabularnewline
\hline 
\hline 
full search & $1024\times550$ & 153.36 & 217.15 & 208.62 & 248.55 & 329.53 & 313.33 & 94.61 & 103.18  & 90.30 \tabularnewline
\hline 
2 & $1024\times50$ & 21.07 & 15.86 & 12.14 & 20.77 & 21.49 & 20.88 & 18.9 & 11.54  & 12.22 \tabularnewline
\hline 
1 & $1024\times30$ & 14.46 & 14.00 & 12.73 & 16.80 & 11.41 & 12.73 & 14.7 & 13.61  & 9.31 \tabularnewline
\hline 
0.5 & $1024\times20$ & 12.88 & 14.15 & 12.09 & 15.61 & 12.09 & 12.67 & 15.9 & 11.28  & 8.91 \tabularnewline
\hline 
0.2 & $1024\times14$ & 13.33 & 14.00 & 11.77 & 15.50 & 14.22 & 12.84 & 16.2 & 11.54  & 9.37 \tabularnewline
\hline 
0.05 & $1024\times12$ & 11.52 & 15.46 & 11.18 & 15.60 & 17.11 & 13.38 & 15.9 & 13.34  & 8.79 \tabularnewline
\hline 
\end{tabular}
\par\end{centering}

\normalsize

\label{table:experimental_results-hill}
\end{table*}

Table~\ref{tab:Local-memory-usage} lists the local memory usage
for different patch sizes for the ``WalkingMan'' test image. The
required local memory for all the patch sizes listed in Table~\ref{tab:Local-memory-usage}
can be fit into the 8KB local memory of the Adreno GPU. In addition,
if we carefully design the method to load data from the global memory
to the local memory by data striping, we can coalesce the global memory
access to further reduce latency. The parallel data loading from the
global memory to the local memory is shown in Algorithm~\ref{alg:Parallel-data-loading},
in which the coalesced global memory access is achieved. 

Experimental results in Fig.~\ref{fig:local_mem_results} demonstrate
the performance improvement by utilizing the local memory to enable
the data sharing between work items inside the same work group. We
observe on average a 30\% reduction in processing time after using
the local memory.

\section{Experimental Results\label{sec:Experimental-Results}}

\begin{figure}
\begin{centering}
\includegraphics[width=1\columnwidth]{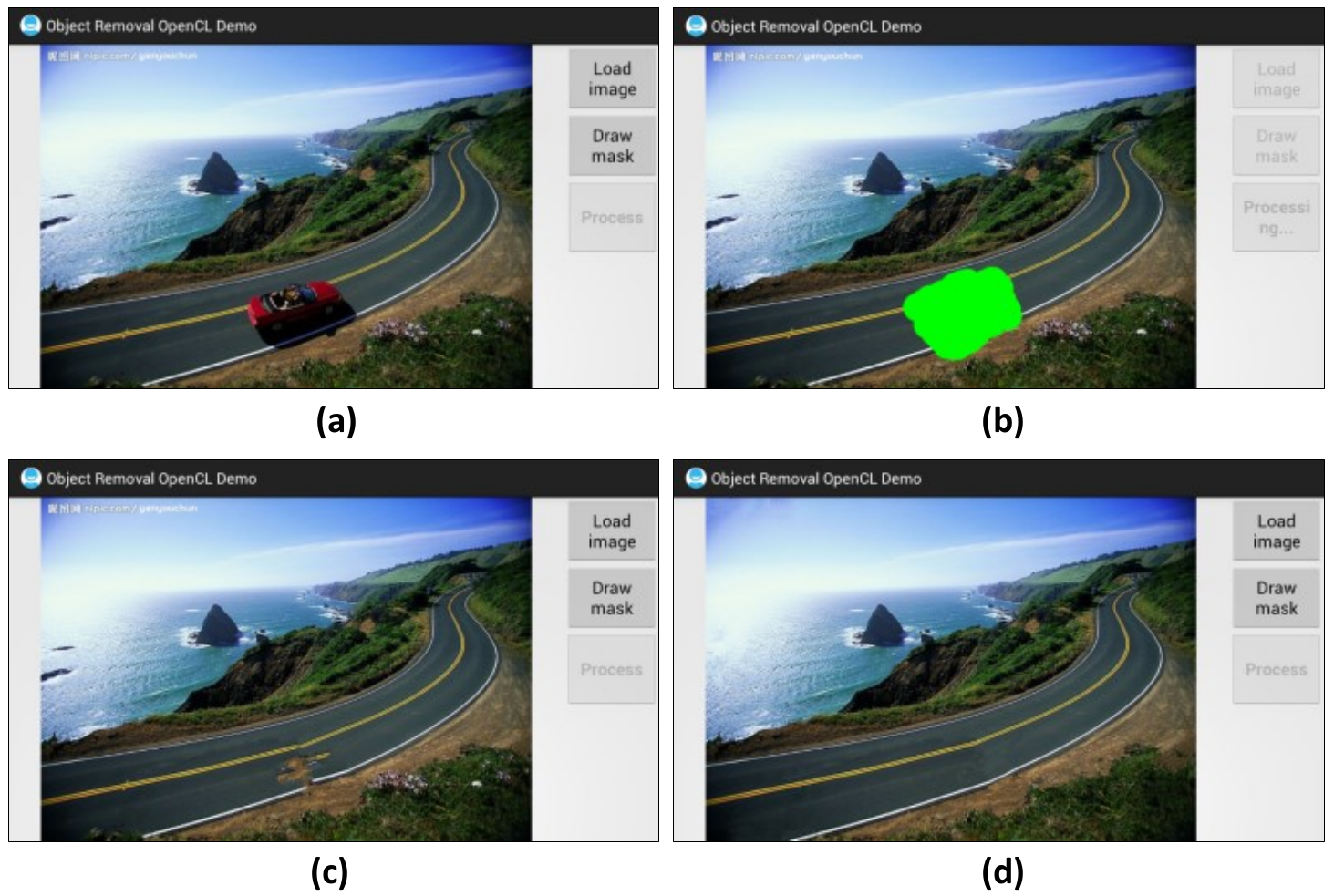}
\par\end{centering}

\caption{An interactive object removal demo on Android with the OpenCL acceleration.
(a) original image; (b) a mask indicating the object area; (c) intermediate
result; (d) final result image after iterative editing.\label{fig:demo}}
\end{figure}

We implemented the exemplar-based inpainting algorithm for object
removal on a test platform based on the Snapdragon chipset using OpenCL
and the Android NDK~\citep{opencl,android}. We applied the proposed
optimization techniques discussed in Section~\ref{sec:Algorithm-Optimizations}.
To demonstrate the efficiency and practicality of the proposed implementation,
we developed an interactive OpenCL Android demo on the test platform.
Fig. \ref{fig:demo} shows screen-shots of the implemented Android
demo application, in which an end user can draw a customized mask
by touching the touchscreen to cover an unwanted object and then remove
it by pressing a button. The demo allows iterative editing, so that
the user can keep editing an image until a satisfying result is obtained.

From Table~\ref{table:experimental_results_walkingman} to Table~\ref{table:experimental_results-hill},
we show the processing time of the OpenCL-based implementations, including
the CPU-only results (utilizing multi-core CPU on the chip) and CPU-GPU
heterogeneous implementations. We can notice that the OpenCL implementations
with proposed optimizations significantly improve the processing performance
compared to the serial version implementation. The CPU-GPU heterogeneous
implementations further improve the performance compared to the CPU-only
implementations. 

Table~\ref{table:experimental_results_walkingman} shows experimental
results for the ``WalkingMan'' image, in which the image size is
$512\times384$, and the size of the object area $76\times128$. The
mask is manually drawn to cover the walking person. With the default
parameter configuration ($9\times9$ patch size, full search), the
serial version of the implementation running on one CPU core uses
87 seconds to finish the processing (shown in Table~\ref{table:breakdown}),
which is a long processing time for a practical mobile application.
The fact that iterative editing is required under many circumstances
makes the serial implementation far from being practical. Table \ref{table:experimental_results_walkingman}
shows experimental results for OpenCL-based parallel solutions. With
the default parameter configuration, the multi-core CPU-only version
reduces the processing time to 23.26 seconds, and the heterogeneous
CPU-GPU implementation further reduces the processing time to 20.37
seconds (76.6\% time reduction compared to 87 seconds processing time
for the serial implementation). 

With all the proposed optimization techniques applied, we observe
significant performance speedup. With search factor $\alpha=0.05$,
patch size $17\times17$, and local memory enabled, the processing
time is reduced to only 1.25 seconds, which indicates a 93.9\% reduction
in processing time compared to 20.37 seconds for the default parameter
configuration (full search, $9\times9$ patch, without using the local
memory). The subjective quality of resultant images does not degrade
according to experimental results shown in Fig.~\ref{fig:effect-reducing-search-area}
and Fig.~\ref{fig:effect-increase-patch-size}. According to the
research conducted by Niida \emph{et al.}, users of mobile applications
can tolerate several seconds average processing time for mobile services
before they start to feel frustrated~\citep{Niida:VehMag2010:waiting_time}.
By accelerating the object removal algorithm using heterogeneous CPU-GPU
partitioning on mobile devices, we successfully reduce the run time,
which makes these types of computer vision algorithms feasible in
practical mobile applications. 

\begin{table}
\scriptsize

\caption{Speedup for OpenCL-based heterogeneous implementations with the proposed
algorithmic and memory optimizations. }

\begin{centering}
\begin{tabular}{c|c|c|c}
\hline 
\multirow{4}{*}{\textbf{Image}} & \multicolumn{2}{c|}{\textbf{Processing time (s)}} & \multirow{4}{*}{Speedup}\tabularnewline
\cline{2-3} 
 & w/o opt. & w/ opt. & \tabularnewline
\cline{2-3} 
 & Full search & $\alpha=0.05$ & \tabularnewline
 & Patch size $9\times9$ & Patch size $17\times17$ & \tabularnewline
\hline 
WalkingMan & 20.37 & 1.25 & 16.3 X\tabularnewline
\hline 
River & 18.28 & 1.00 & 18.3 X\tabularnewline
\hline 
DivingMan & 44.83 & 5.31 & 8.44 X\tabularnewline
\hline 
Hill & 248.55 & 8.79 & 28.3 X\tabularnewline
\hline 
\end{tabular}
\par\end{centering}

\normalsize\label{table:speedup}
\end{table}

We can draw similar conclusions from the timing results for other
test images shown in Table~\ref{table:experimental_results-river},
\ref{table:experimental_results-dive} and \ref{table:experimental_results-hill}.
To demonstrate the effectiveness of our proposed optimization schemes,
the speedup gained from the proposed optimization strategies are concluded
in Table~\ref{table:speedup}. We observe speedups from 8.44X to
28.3X with all our proposed optimizations applied to the heterogeneous
OpenCL implementations.

\section{Conclusions\label{sec:Conclusions}}

The emerging heterogeneous architecture of mobile SoC processors with
the support of parallel programming models such as OpenCL enables
the capability of general-purpose computing on the mobile GPU. As
a case study, we present an OpenCL-based mobile GPU implementation
of an object removal algorithm. We analyze the workload for the computationally-intensive
kernels of the algorithm, and partition the algorithm between the
mobile CPU and GPU. Algorithm mapping methodology and optimization
techniques for OpenCL-based parallel implementation are discussed.
Several implementation trade-offs are discussed according to the architectural
properties of the mobile GPU. We perform experiments on a real mobile
platform powered by a Snapdragon mobile SoC. The experimental results
show that the CPU-GPU heterogeneous implementation reduces the processing
time to 20.37 seconds, compared to 87 seconds processing time for
the single-thread CPU-only version. With the proposed optimization
strategies, the processing time can be further reduced without degrading
the visual image quality. When we apply the proposed optimizations
(setting the search factor to $\alpha=0.05$ and increasing the patch
size to $17\times17$), we observe speedups from 8.44X to 28.3X for
different test images. With the rapid development of mobile SoCs with
heterogeneous computing capability, we foresee that many more computer
vision applications with high complexity will be enabled on real-world
mobile devices.

\section*{Acknowledgment}

This work was supported in part by Qualcomm, and by the US National
Science Foundation under grants CNS-1265332, ECCS-1232274, and EECS-0925942.

\small

\bibliographystyle{spbasic}

\end{document}